\documentclass[review]{elsarticle}

\usepackage{amssymb}
\usepackage{amsmath,bm}
\usepackage{float}
\usepackage{algorithm}
\usepackage{algpseudocode}
\usepackage{booktabs}
\usepackage{graphicx}
\usepackage{multirow}
\usepackage{xcolor}
\usepackage{bm}
\usepackage{colortbl}
\usepackage{setspace}
\usepackage{natbib} 
\biboptions{sort&compress}
\usepackage{arydshln}
\usepackage[hidelinks,colorlinks=true,linkcolor=blue]{hyperref}
\usepackage{graphicx}%
\usepackage{multirow}%
\usepackage{amssymb,amsfonts}%
\usepackage{amsthm}%
\usepackage{mathrsfs}%
\usepackage[title]{appendix}%
\usepackage{textcomp}%
\usepackage{manyfoot}%
\usepackage{booktabs}%
\usepackage{listings}%
\usepackage{colortbl}
\usepackage[right=1in, left=1in, top=1in, bottom=1in]{geometry} 
\usepackage[font=normalsize,labelfont=bf]{caption}
\usepackage{subcaption}
\makeatletter
\def\ps@pprintTitle{%
 \let\@oddhead\@empty
 \let\@evenhead\@empty
 \def\@oddfoot{\reset@font\hfil\thepage\hfil}
 \let\@evenfoot\@oddfoot
}
\makeatother

\begin{document}

\begin{frontmatter}

\title{Adaptive Loss Weighting for Machine Learning Interatomic Potentials}

\author[1]{Daniel Ocampo}
\author[2]{Daniela Posso}
\author[1]{Reza Namakian}
\author[1,3]{Wei Gao\corref{correspondingauthor}}
\ead{wei.gao@tamu.edu}

\address[1]{J. Mike Walker $'$66 Department of Mechanical Engineering, Texas A\&M University, College Station, Texas 77843, United States}
\address[2]{Department of Mechanical Engineering, The University of Texas at San Antonio, San Antonio, Texas 78249, United States}
\address[3]{Department of Materials Science \& Engineering, Texas A\&M University, College Station, Texas 77843, United States}
\cortext[correspondingauthor]{Corresponding author}

\begin{abstract}

Training machine learning interatomic potentials often requires optimizing a loss function composed of three variables: potential energies, forces, and stress. The contribution of each variable to the total loss is typically weighted using fixed coefficients. Identifying these coefficients usually relies on iterative or heuristic methods, which may yield sub-optimal
 results. To address this issue, we propose an adaptive loss weighting algorithm that automatically adjusts the loss weights of these variables during the training of potentials, dynamically adapting to the characteristics of the training dataset. The comparative analysis of models trained with fixed and adaptive loss weights demonstrates that the adaptive method not only achieves a more balanced predictions across the three variables but also improves overall prediction accuracy.

\end{abstract}


\end{frontmatter}

\section{Introduction}\label{introduction}

Machine learning inter-atomic potentials (ML-IAPs) can be broadly split into two types. The first is descriptor-based ML-IAP, in which the descriptors (or fingerprints) are used to describe the environment of the atoms in a system. Various descriptors have been proposed in the literature, such as Atom-Centered Symmetry Functions (ACSF) \cite{behler2007generalized}, Smooth Overlap of Atomic Positions (SOAP), Atomic Cluster Expansion (ACE) \cite{drautz2019atomic}, and Moment Tensor Potentials \cite{shapeev2016moment}, among others. A comprehensive review of the descriptors can be found in Musil et al.'s work \cite{musil2021physics}.  Representative descriptor-based ML-IAPs include: Behler and Parrinello Neural Network potential \cite{behler2007generalized, behler2011atom}, Gaussian approximation potential (GAP) \cite{bartok2010gaussian}, Spectral Neighbor Analysis Potential (SNAP) \cite{thompson2015spectral}, Moment Tensor Potential (MTP) \cite{shapeev2016moment}, Performant implementation of the atomic cluster expansion (PACE) \cite{lysogorskiy2021performant}, and DeePMD \cite{wang2018deepmd} among others. 
The second type of ML-IAP is the end-to-end potential, which operates differently by learning directly from the types and positions of atoms, without the need for predefined descriptors.  Representative ones include: Crystal Graph Convolutional Neural Networks (CGCNN) \cite{xie2018crystal, park2020developing}, SchNet \cite{schutt2018schnet}, and MatErials Graph Network (MEGNet) \cite{chen2019graph} among others. Although the end-to-end ML-IAPs leverage more recent and advanced feature learning AI technology, there is currently no conclusive evidence to suggest that end-to-end ML-IAPs outperform the descriptor-based ML-IAP in terms of prediction accuracy. The study reported in this paper are performed using ACSF.

The training of most ML-IAPs involves minimizing a loss function, which measures the difference between the predicted outputs of the potential and the actual target value obtained from Ab initio simulations. Typically, an ML-IAP's loss function comprises three components: potential energy, atomic forces, and stress tensor, each weighted by a prefactor (i.e. loss weight). Most of current ML-IAPs assign a predefined loss weight to each component, which stays as a constant throughout the training process \cite{behler2011atom, thompson2015spectral, shapeev2016moment, lysogorskiy2021performant}. The approaches like DeepMD modulate these weights linearly during training, though the rationale and effectiveness of this approach are not fully clear \cite{wang2018deepmd, zeng2023deepmd}.  In this paper, our study demonstrates that varying combinations of loss weights significantly impact model performance, raising a key question: what constitutes an effective combination of loss weights that balances energy, force, and stress predictions?  Our hypothesis is that a universally optimal set of loss weights for all ML-IAPs may not exist, as the optimal weights are likely tied to the material system and the characteristics of individual training datasets. Instead, we propose a new method that automatically adjusts the loss weights during the training of potentials. This approach dynamically adapts to the characteristics of the training dataset and optimizes ML-IAPs predictions.

The paper is structured as follows: first, we provide an overview of ML-IAPs based on neural networks, including the formulation of loss functions. Next, we introduce the principle and algorithm of adaptive loss weighting. Then, the results from models using adaptive methods are compared with those from models using fixed loss weights, in order to demonstrate the advantage of the adaptive method. Finally, the paper is concluded with a summary of main findings.

% =================================== METHODS =================================== 
\section{Computation Methods}\label{computation_methods} 

The adaptive loss weighting algorithm can be applied to a wide range of ML-IAPs that require minimizing a loss function. In this study, to demonstrate its applicability, we implemented it within the framework of the Behler and Parrinello type of neural network potential \cite{behler2007generalized, behler2011atom} inside an open-source  package AtomDNN \cite{gao2021atomdnn} developed by the authors.

\subsection{Neural Network Potentials}

\begin{figure}[t!]
    \centering
    \includegraphics[width=0.9\textwidth]{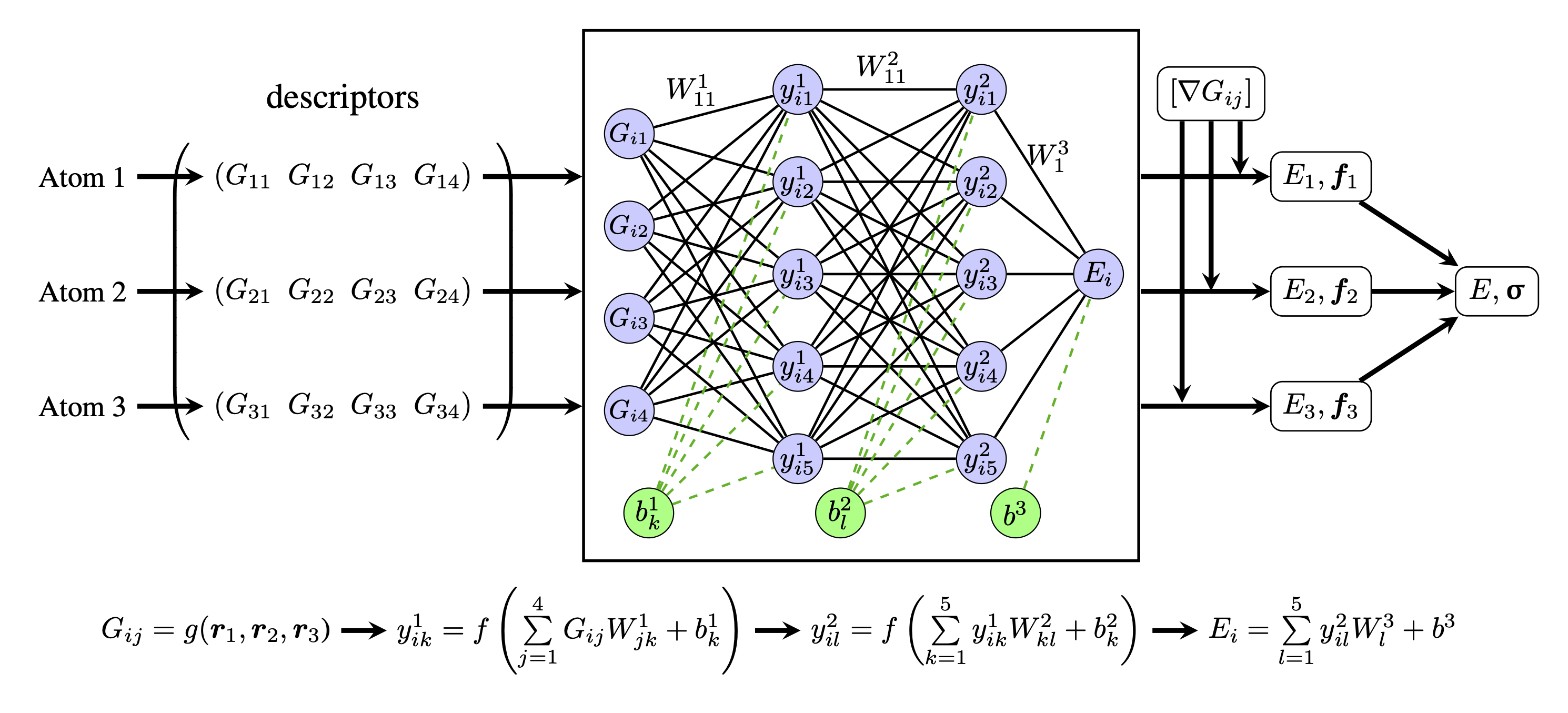}
    \captionsetup{width=.9\linewidth}
    \caption{Schematic of neural network interatomic potential where the potential energy, forces, and stresses are calculated using a feed-forward, descriptor-based neural network.}
    \label{fig:Diagram-Force-and-stress}
\end{figure}

In a descriptor-based neural network potential designed for a system with $N$ atoms, the Cartesian coordinates of each atom $i$ are denoted as $\bm{r}_i=\{r^{(1)}, ..., r^{(N)}\}$, which are transformed into a descriptor vector $\bm{G}_{ij}$ with $M$ components, where $j=1,...,M$. While many descriptor types can be integrated with neural networks, in this study, atom-centered symmetry functions (ACSF) are utilized.   
The working principle of the neural network potential can be explained in Fig. \ref{fig:Diagram-Force-and-stress}, where a scenario with three atoms of the same chemical species is considered. These atoms are represented by 4 component descriptors, which feed into the neural network as inputs.
This network consists of two densely connected layers, each containing 5 neurons, and concludes with a linear concatenation layer. The network can be considered as a high dimensional non-linear function mapping, parametrized with weight matrices and bias vectors which are optimized during training. The equations in the figure constitute a forward-pass through the network, where $W^1_{jk}$, $W^2_{kl}$ and $W^3_{l}$ are the weight matrices of layers 1, 2 and, 3, respectively, $b^1_k$, $b^2_l$, and $b^3$ are the corresponding biases, and $f$ is the activation function. 

In this model, each atom of the same chemical species is processed through an identical neural network, yielding a per-atom energy $E_i$. In systems comprising multiple chemical species, separate parallel networks are employed for each species. The total potential energy of the system, $\cal E$, is the sum of per-atom energies, which can be written in terms of descriptors as
\begin{equation}
{\cal E}= \sum_i^N E_i = \sum_i^N E_{i}(G_{i1},G_{i2},...,G_{iM}), \label{eq:energy}
\end{equation}
where $i$ is the index for individual atom that is described by a feature vector with $M$ components. The atomic forces can be obtained from the negative gradients of the total potential energy with respect to the atomic positions, which can be expressed in terms of the derivatives of descriptors using a chain rule
\begin{equation}
f_{j\alpha} = -\frac{\partial {\cal E}}{\partial r_{j\alpha}} = -\sum_{i=1}^{N}\sum_{m=1}^{M}\frac{\partial E_{i}}{\partial G_{im}}\frac{\partial G_{im}}{\partial r_{j\alpha}},\label{eq:force-dEdr}
\end{equation}
where  $r_{j\alpha}(\alpha=1,2,3)$ are the Cartesian coordinates
of the \emph{j}-th atom. The derivatives of descriptors, $\partial G_{im}/\partial r_{j\alpha}$, are fed to the network as additional inputs (represented by $\nabla G_{ij}$ in Fig. \ref{fig:Diagram-Force-and-stress}). The other derivative term, $\partial E_{i}/\partial G_{im}$, is readily available from the back-propagation algorithm used during the training process.  In addition, for solid-state materials, Cauchy stress tensor can be used for training ML-IAPs, which can be written as 
\begin{equation} \label{eq: stress}
	\sigma_{\alpha\beta} = \frac{1}{V}  \sum_{i=1}^{N} \sum_{m=1}^{M} \sum_{j \in \text{NB}_i} \frac{\partial E_i}{\partial G_{im}} \frac{\partial G_{im} } {\partial r_{j \alpha}} r_{j\beta},
\end{equation}
where $V$ is the current volume and $\text{NB}_i$ represents the neighbor list of atom $i$ withing the cutoff distance $r_c$. The derivation of Eq. (\ref{eq: stress}) can be found in \ref{sec: appendix a}.

\subsection{Loss Function of ML-IAPs} \label{sec: loss_function}
The loss function for training neural network potential as well as many other types of ML-IAPs can be written as Eq. (\ref{eq:loss_function}), which is composed of three components: potential energy, forces, and stresses.
\begin{equation}\label{eq:loss_function}
    \mathcal{L} = 
    \frac{\alpha_1}{N}  \sum_{i=1}^N\left(\hat{E}_i  - E_i\right)^2 +
    \frac{\alpha_2}{3N} \sum_{i=1}^N \sum_{\alpha=0}^3 \left(\hat{f}_{i\alpha} - f_{i\alpha}\right)^2 +
    \frac{\alpha_3}{6} \sum_{j=1}^6 \left(\hat{\sigma}_{j}  - \sigma_{j}\right)^2,
\end{equation}
where $\alpha_1$, $\alpha_2$, and $\alpha_3$ are the weight coefficients to balance each component towards the calculation of the total loss. The variables with hat refer to the true values from DFT calculations, and the Voigt notation is applied to stress components. 
Previous studies have indicated that training of ML-IAP with both energy and force is beneficial as it  integrates data from potential energies and their gradients, leading to more stable predictions and a reduction in the quantity of data structures needed for effective training \cite{pukrittayakamee2009simultaneous, cooper2020efficient}. The loss term for stresses has also been shown to be beneficial in promoting transferability of the ML-IAP \cite{yanxon2020developments}. However, the specific impact of incorporating stress in the loss has not been thoroughly investigated. It is also noteworthy that most prior studies have limited their focus to only the potential energy and force components in training, which could lead to unsatisfactory stress predictions, as our later computation results will show.

% There is also a paper by Yanxon, using lw={1, 1e-3, 1e-4} \cite{yanxon2020developments}

In previous studies, weight coefficients for ML-IAPs were typically kept constant during training. For instance, in a recent study, a series of  ACE-based potentials were trained using various energy and force weight combinations on a copper dataset. The findings suggested that an $0.8:0.2$ ratio of potential energy to force loss weight was optimal for performance \cite{bochkarev2022efficient}. In anther study, the ACE-based potential was trained by assigning varied loss weight coefficients to different subsets of the training data, though the methodology for selecting these parameters was not detailed \cite{lysogorskiy2021performant}. Moreover, instead of fixed weight coefficients, a different approach was taken in training a neural network potential using DeepMD-kit. This method uses a monotonic decrease in the force weight while increasing the energy weight during the training of a dataset containing pure Silica zeolite structures \cite{sours2023predicting}. However, the effectiveness of this particular training strategy was not detailed.

\subsection{Adaptive Loss Weighting\label{sec:adaptive_loss}}

In this study, we propose to dynamically adjust loss weights during the potential training process. This concept is inspired by the work of Heydari et al. \cite{heydari2019softadapt}, who developed an adaptive loss weighting algorithm for a convolutional neural network for image reconstruction and synthetic data generation. Similar to their strategy, our method was designed to manage multipart loss functions by dynamically altering each loss weight. This is accomplished by continuously recalibrating the loss weights based on the change in loss values from one epoch to the next, ensuring a balanced optimization process without any single component becoming overly dominant. The advantages of this approach are twofold: it avoids the scenarios for certain loss components to disproportionately impact the training process and eliminates the need for manually determining the optimal fixed weight combination each time the material system or dataset changes.

\begin{algorithm}[t!]
\caption{Adaptive Loss Weighting for gradient descent based neural network training. The set of trainable variables of the model are represented by $\theta$. Note, batching is omitted here for simplicity purposes. Normalization of $\textbf{s}$  vector is considered as a default for reasons previously mentioned.\label{alg:softadapt_algorithm}}
\begin{algorithmic}[1]
\begin{spacing}{1.}
\Require $optimizer$
\Require $loss\_fn$ (loss function, which calculates the difference between target $y_k$ value and prediction $h(x,\theta)$)
\Require $n$ (update loss weights every $n$ epochs)
\Require $\alpha_k^{(0)}$ (initial loss weights values)
\Require $\epsilon=10^{-8}$ for numerical stability

\State $loss\_weights \gets list()$ empty list to store average loss weights for $n$ epochs
\State $\alpha_k^{(i)} \gets \alpha_k^{(0)}$    

\For{$i = 1$ \textbf{in} $epochs$}
    % \State $l_k^{(i)} \gets$  $loss\_fn(\textbf{y}, NN(\textbf{x}, \textbf{\theta}))$ Compute loss for current loss weights
    \State $l_k^{(i)} \gets loss\_fn(\textbf{y}, h(\textbf{x}, \bm{\theta}))$ Compute loss for current loss weights
    \State $l_T \gets \alpha_k^{(i)}$ Compute total loss using current loss weights
    \State Perform back-propagation to update $\theta$ parameters

    \If{$(epoch \% n) == 0$}
        \State $epoch\_loss\_weights \gets list()$ empty list to store loss weights for current epoch
    \EndIf

    \State $s_k^{(i)} \gets l_k^{(i)} - l_k^{(i-1)}/ \left((\sum_{m=1}^{M}|s|)+\epsilon\right)$
    
    \State $\alpha_k^{(i)} \gets \exp(\beta s_k^{(i)}) / \left((\sum_{m=1}^M \exp(\beta s_m^{(i)})) + \epsilon\right)$
    
    \State Append $\alpha_k^{(i)}$ to $epoch\_loss\_weights$

    \If{$[(epoch + 1) \% n] == 0$}
        \State $loss\_weights \gets$ Compute average of $n$ entries in $epoch\_loss\_weights$
    \EndIf
\EndFor
\end{spacing}
\end{algorithmic}
\end{algorithm}

The proposed adaptive method is presented in Algorithm \ref{alg:softadapt_algorithm}. To illustrate the machinery of the algorithm, we first re-write Eq. (\ref{eq:loss_function}) as 
\begin{equation}
    \mathcal{L}_i=\sum_{k=1}^3\alpha_k^i\ell_k , \label{eq:loss_function2}
\end{equation}
where $\ell_1$, $\ell_2$, and $\ell_3$ are the loss components for potential energy, forces, and stresses, respectively. The weight factor $\alpha_k$ is calculated at the $i$-th epoch by 
\begin{equation}
\alpha_{k}^{i}=\frac{e^{\beta\cdot s_{k}^{i}}} {\sum_{m=1}^{3}e^{\beta\cdot s_{m}^{i}}},\label{eq:alpha}
\end{equation}
where
\begin{equation}
    s^{i}_k = \ell_{k}^{i} - \ell_{k}^{i-1}
\end{equation}
represents the rate of change of $k$-th loss component, scaled by a hyperparameter $\beta$. Eq. (\ref{eq:alpha}) adopts the classic Softmax function, and therefore, the algorithm is referred to as \textit{Softadapt}. This equation indicates that a positive value of $\beta$ places a higher weight on the component with the most positive rate of change, which corresponds to the worst performing component in terms of learning. On the other hand, a negative value of $\beta$ results in a higher relative weight for the term with the most negative rate of change or the best performing component. A zero value of $\beta$ yields equal fixed weights for each loss component. The magnitude of $\beta$ determines how sensitive the loss weights respond to the changes in individual loss components, with larger $\beta$ values leading to more responsive adjustment. While the Softadapt method may require adjusting the hyperparameter $\beta$, it offers greater convenience compared to the iterative adjustment of the ratio among three loss weights required in the fixed loss weight approach.

\subsection{Benchmark Data Generation}

Due to lack of accessible datasets comprising high-fidelity stress data, we created a benchmark dataset using a two-dimensional Molybdenum Ditelluride (MoTe$_2$) as a model material. The dataset consists of a total of $3,146$ structures, evenly distributed between the 2H and 1T' phases of  MoTe$_2$. These structures are subjected to various tensile and compressive strains, both uniaxial and biaxial, ranging from $-10\%$ to $10\%$. The data include both the deformed equilibrium structures and the structures in which atoms are random perturbated, i.e., randomly drawn from a normal distribution with $0.1$ {\AA} standard deviation.

All DFT calculations for data generation were performed using the plane-wave-based Vienna Ab-initio Simulation
Package (VASP) \cite{kresse1996efficient, kresse1993ab}. Projector augmented wave (PAW) pseudopotentials \cite{kresse1999ultrasoft, blochl1994projector} were used to represent ionic cores, and the electronic kinetic energy cutoff for the plane-wave basis describing the valence electrons was set to 293 eV. The Perdew-Burke-Ernzerhof (PBE) with the generalized gradient approximation (GGA) \cite{perdew1996generalized} was chosen for the exchange-correlation functional. The k-point selection was adapted to the dimension of each structure along the armchair and zigzag directions, keeping a k-point resolved value of $0.02$$\pi\times\text{\r{A}}^{-1}$ in both directions, following the Monkhorst-Pack scheme \cite{monkhorst1976special} in VASPKIT \cite{wang2021vaspkit}. As for the out-of-plane direction, a vacuum layer of $25$ {\AA} was used to separate the periodic images in the out-of-plane direction, thereby single k-point was maintained along this direction. The electronic energy and atomic forces were converged to $10^{-4}$ meV and $1$ meV/\r{A}, respectively.

\section{Results and Discussion}\label{results}

The neural network models in this study are designed with a specific architecture, consisting of two hidden layers, each containing 30 neurons. A hyperbolic tangent function serves as the activation mechanism. The optimization is performed using the Adam optimizer, set at a learning rate of 0.001. Training of these models continued until negligible learning gains were observed. The dataset was divided into 70\% training, 20\% validation, and 10\% testing.

\subsection{Results of Fixed Loss Weights}\label{subsec:results_fixed_lw}

As dicussed in Section \ref{sec: loss_function}, while various weighting schemes have been used in the literature, a detailed study that simultaneously considers potential energy, forces, and stresses in the loss function and investigates their weightings' impacts on the training performance is still missing. To address this, we first conducted a sequence of studies focusing on the effects of fixed loss weights. This involved systematically varying these weights and analyzing their impact on the performance of the trained model in terms of predictive capabilities in comparison with DFT calculations. The results are shown in Table~\ref{tab:fixed_lw_study}.

\begin{table}[t!]
\begin{tabular*}{\textwidth}{@{\extracolsep\fill}cccccccccc}
\toprule%
 & \multicolumn{3}{>{\columncolor[gray]{0.8}}c}{Fixed Loss Weights}
 & \multicolumn{3}{>{\columncolor[gray]{0.8}}c}{Training RMSE}
 & \multicolumn{3}{>{\columncolor[gray]{0.8}}c}{Testing RMSE}\\

 Model & $\alpha_1$ & $\alpha_2$ & $\alpha_3$ & Energy & Force & Stress & Energy & Force & Stress\\
 \cmidrule{1-1}\cmidrule{2-4}\cmidrule{5-7}\cmidrule{8-10}%
1 & 1.0 & 0.00 & 0.00 & 2.37 & 1934.61 & 2832.32 & 19.28 & 2940.90 & 4890.23\tabularnewline
2 & 0.90 & 0.10 & 0.00 & 1.98 & 15.87 & 489.98 & 2.30 & 24.28 & 520.72\tabularnewline
3 & \textbf{0.90} & 0.05 & 0.05 & 2.72 & 21.39 &  22.02 & \textbf{2.29} & 28.42 & 34.10\tabularnewline
4 & 0.05 & \textbf{0.90} & 0.05 & 7.61 & 16.00 &  30.17 & 6.12 & \textbf{21.39} & 37.20\tabularnewline
5 & 0.05 & 0.05 & \textbf{0.90} & 7.79 & 42.69 &  15.48 & 7.42 & 48.43 & \textbf{23.43}\tabularnewline
6 & 0.33 & 0.33 & 0.33 & 3.58 & 20.23 &  18.11 & 5.41 & 25.50 & 22.44\tabularnewline
\bottomrule
\end{tabular*}
\captionsetup{width=.9\textwidth}
\caption{RMSE for models trained with different loss weights combinations. The contribution
of potential energy (meV/atom), force (meV/\text{\AA}), and stress (MPa) are weighted by $\alpha_1$, $\alpha_2$, and $\alpha_3$, respectively. 
\label{tab:fixed_lw_study}}
\end{table}

In model 1, where the loss function incorporates only potential energy, the model's predictions for the forces and stresses are notably poor.  Additionally, the testing error for energy in this model is an order of magnitude higher than that of other models despite favorable training error. This suggests that the model trained with energy alone lacks generalization capabilities. In the case of model 2, which is trained on both potential energy and atomic forces, we observe good performance in predicting energy and forces. However, this approach leads to inaccurate stress predictions. Apparently, the absence of stress data during training notably deteriorates the model's ability to predict stress accurately. Although this was observed in this particular dataset, it is likely applicable to others as well. Model 3, which is trained on all three components delivers more balanced results across all metrics.  The comparison between model 2 and 3 emphasizes the importance of incorporating stress data in training ML-IAPs to more accurately predict the stresses, in addition to energies and forces.

Models 3 to 5 were each trained with distinct weight combinations, specifically designed to prioritize one of three variables: potential energy, force, or stress.  The results showed that the weight coefficient for one variable improves its corresponding result but at the cost of reducing accuracy in the other one or two variables. Specifically, model 3, which prioritized potential energy, exhibited the lowest testing loss for potential energy but did so by sacrificing stress prediction accuracy. Model 4, focusing on force, achieved the lowest force testing loss, but compromising energy and stress accuracy. Similarly, model 5 achieved the best results for stress at the expense of potential energy and force accuracy. This observed trade-off pattern suggests that previously reported approaches, which favor the forces during potential training (such as $\alpha_1:\alpha_2=1:10$) \cite{singraber2019parallel}, or completely overlook the energy term and focus exclusively on forces \cite{chmiela2018towards}, might not be universally applicable across different datasets or material systems. Additionally, model 6, which distributed weights evenly across all three variables, managed to achieve satisfactory test results for force and stress, yet it did not accurately predict the potential energy. Therefore, this fixed weight approach could require extensive experimentation with loss weight combinations, specific to the material dataset, in order to attain a balanced performance among all three variables. This suggests the potential benefits of employing automated methods to streamline the training process.

\subsection{Results using Adaptive Loss Weights \label{subsec:results_adaptive}}

As discussed in the previous section, achieving optimal results for potential energy, force, and stress simultaneously may require iteratively adjusting the loss weight ratios based on material and dataset characteristics. By contrast, we propose the Softadapt method to dynamically balance each term's contribution to the loss function. In this section, the results of models trained using Softadapt are compared with those trained using fixed weights. For a fair comparison, all neural network models implementing the adaptive algorithm were configured with the same architecture and activation function to their fixed weight counterparts.

\begin{table}[b!]
\begin{tabular*}{\textwidth}{@{\extracolsep\fill}ccccccccc}
\toprule 
     & \multicolumn{1}{>{\columncolor[gray]{0.8}}c}{Softadapt}
     & \multicolumn{3}{>{\columncolor[gray]{0.8}}c}{Training RMSE}
     & \multicolumn{3}{>{\columncolor[gray]{0.8}}c}{Testing RMSE}\\
    Model &  $\beta$ & Energy & Force & Stress & Energy & Force & Stress\tabularnewline
    \midrule 
    1 &0.01	& 4.58 & 26.51 & 25.67 & 4.03 & 32.52 & 29.70\tabularnewline
    2 &	0.1		& 3.47 & 21.20 & 19.42 & 4.15 & 27.27 & 27.08\tabularnewline
    3 &1.0		& 3.07 & 18.32 & 14.55 & 2.74 & 23.85 & 23.54\tabularnewline
    4 &	2.0		& 2.80 & 17.90 & 15.24 & 2.57 & 25.70 & 24.62\tabularnewline
\bottomrule
\end{tabular*}
\captionsetup{width=.9\textwidth}
\caption{RMSE for models trained using Softadapt algorithm with different $\beta$ values. Units for potential energy, force, and stress are meV/atom, meV/\text{\AA}, and MPa, respectively. \label{tab:adaptive_lw_study}}
\end{table}

\begin{figure}[t!]
    \centering
    \includegraphics[width=0.5\linewidth]{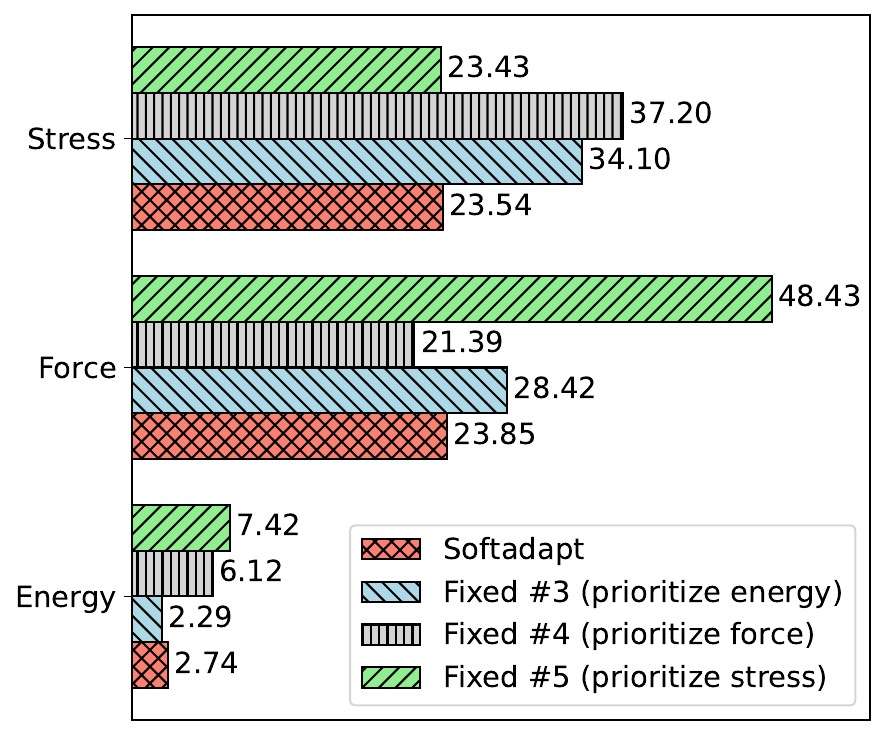}
    \captionsetup{width=.9\linewidth}
    \caption{RMSE comparison between the model trained using Softadpt algorithm and three models trained with fixed loss weights, each prioritizing potential energy, force, and stress. The RMSE values are taken from Table \ref{tab:fixed_lw_study} and \ref{tab:adaptive_lw_study}. Units for potential energy, force, and stress are meV/atom, meV/\text{\AA}, and MPa, respectively.}
    \label{fig:bar_plot}
\end{figure}

The results of models trained with Softadapt algorithm are presented in Table \ref{tab:adaptive_lw_study}. The hyperparameter $\beta$ controls how sensitively the weights respond to rate changes in loss components, with higher $\beta$ values causing more rapid adjustments in loss weights during training. For the benchmark data, the optimal performance of Softadapt algorithm is achieved when $\beta$ is around the order of 1.0, offering a better and more balanced testing errors across energy, force, and stress compared to the fixed loss weight models. When $\beta$ is too low, the weight adjustments don't adequately keep up with loss rate changes. Conversely, when $\beta$ is set too high, it leads to significant fluctuations in loss weights, resulting in unstable training and increased errors. The comparison between models trained using Softadapt algorithm with $\beta = 1$ and those trained using fixed weights is illustrated in Fig.~\ref{fig:bar_plot}. Notably, for each term of potential energy, force, and stress, the Softadapt method achieves an accuracy level equivalent to that of the corresponding fixed model where that specific term is prioritized.

\begin{figure}[b!]
    \centering
    \includegraphics[width=0.9\linewidth]{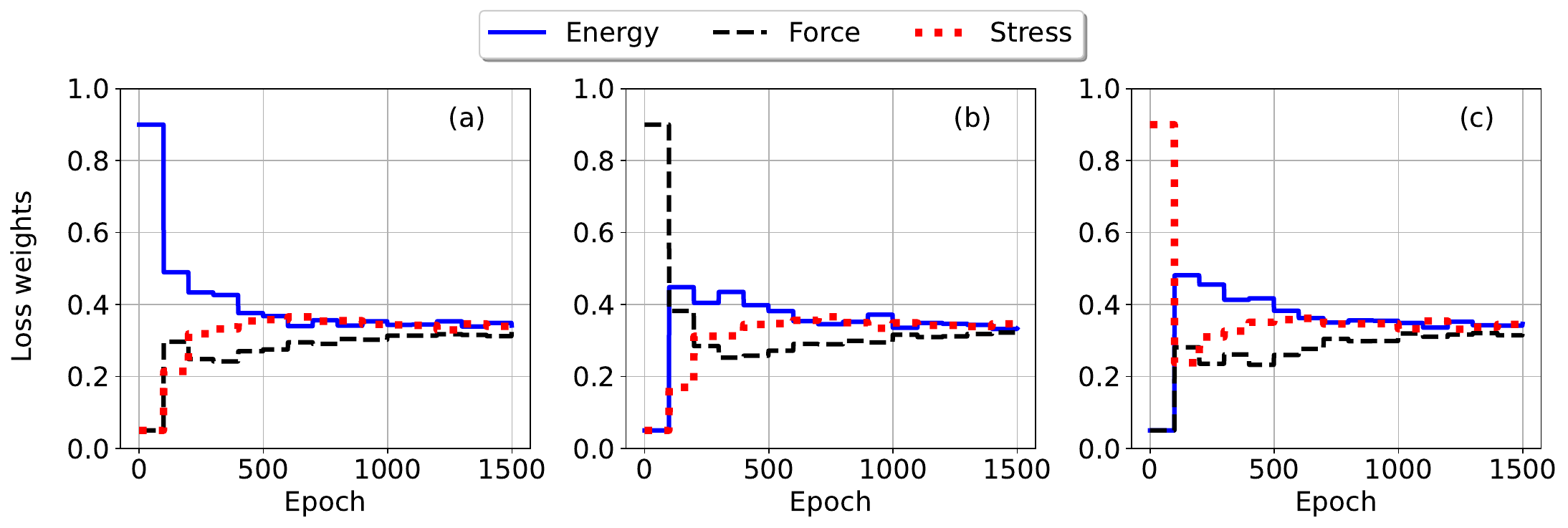}
    \captionsetup{width=.9\linewidth}
    \caption{Variation of loss weights during training using the Softadapt algorithm, with different initial weight ratios among potential energy, force and stress: (a) 0.9:0.05:0.05, (b) 0.05:0.9:0.05 and (c) 0.05:0.05:0.9.}
    \label{fig:loss_weights}
\end{figure}

The results in Table \ref{tab:adaptive_lw_study} were trained using equal initial weights among potential energy, force, and stress, each equals to 0.33. It is important to note that the final results are not sensitive to the selection of the initial weights, due to the nature of adaptive algorithm.  To examine the influence of adaptive algorithm on the changing of loss weights, we monitored the variation of loss weights throughout the training process. In this analysis, we intentionally varied the initial loss weights across three configurations: (0.9, 0.05, 0.05), (0.05, 0.9, 0.05), and (0.05, 0.05, 0.9). As shown in Fig.~\ref{fig:loss_weights}, despite starting with very different initial loss weights, the algorithm adjusts the contribution of each, converging to an optimum combination. Interestingly, for this particular dataset, noticeable adjustment in loss weights mainly occur within the first 1500 epochs. 

\section{Summary}\label{sec:summary}

We proposed an adaptive loss weighting method to dynamically adjust the contribution of penitential energy, force, and stress, based on their corresponding loss values during the training of machine learning interatomic potentials.
Leveraging a benchmark dataset, we conducted a comparative analysis between models trained with fixed and adaptive loss weights. The key findings are summarized as follows.

\begin{itemize}
    \item Stress data proves critical for the accurate prediction of stress values when training machine learning potentials.
    \item Models employing fixed loss weights yield imbalanced predictions for potential energy, force, and stress, as they enhance the accuracy of a prioritized variable at the expense of others.
    \item Models utilizing the adaptive algorithm demonstrate an ability to balance the contributions of the three variables, thereby yielding more balanced and accurate predictions.
\end{itemize}

\section*{Data availability statement}
The data and code reported in this paper are available on Github \cite{gao2021atomdnn}.

\section*{CRediT authorship contribution statement}
\textbf{Daniel Ocampo}: Methodology, Software, Investigation, Writing - original draft, Writing - review \& editing. \textbf{Daniela Posso}: Software, Investigation. \textbf{Reza Namakian}: Investigation, Writing - review \& editing. \textbf{Wei Gao}: Conceptualization, Methodology, Software, Investigation, Writing - original draft, Writing - review \& editing, Supervision, Funding acquisition.

\section*{Acknowledgments}
W.G. gratefully acknowledges financial support of this work by the National Science Foundation through Grant no.
CMMI-2308163 and CMMI-2305529. The authors acknowledge the Texas Advanced Computing Center (TACC) at the University of Texas at Austin and Texas A\&M High Performance Research Computing for providing HPC resources that have contributed to the research results reported within this paper.


\appendix
\section{Cauchy Stress Derivation in ML-IAP} \label{sec: appendix a}

The first Piola-Kirchhoff (PK) stress tensor can be calculated as the work conjugate of deformation gradient tensor
\begin{equation}
	P_{\alpha\beta} = \frac{1}{V_0} \frac{\partial {\cal E}}{\partial F_{\alpha\beta}},
\end{equation}
where $V_0$ is the volume of the reference configuration, and $F_{\alpha\beta}$ is the deformation gradient tensor. Similar to the atomic force calculation, the potential energy can be written as the sum of per-atom potential energies, so the stress can be written as
\begin{equation}\label{eq: pk stress 1}
	P_{\alpha\beta} = \frac{1}{V_0} \sum_{i=1}^{N} \sum_{m=1}^{M} \frac{\partial E_i}{\partial G_{im}} \frac{\partial G_{im} } {\partial F_{\alpha \beta}}.
\end{equation}
The fingerprint $G_{im}$ is determined by the coordinates of the atoms inside the neighbor list of atom $i$. Therefore, using the chain rule, we have
\begin{equation}\label{eq: dGdF}
\frac{\partial G_{im} } {\partial F_{\alpha \beta}} = \sum_{j \in \text{NB}_{i}} \sum_{\gamma=1}^{3} \frac{\partial G_{im}}{\partial r_{j\gamma}} \frac{\partial r_{j\gamma}}{\partial F_{\alpha \beta}},
\end{equation}
where atom $j$ is inside the neighbor list of atom $i$ (represented by $\text {NB}_i$). By definition, the deformation gradient maps the atom position from the reference configuration to the current configuration
\begin{equation}\label{eq: coordinate mapping}
	r_{j\gamma} = \sum_{\beta=1}^{3}F_{\gamma\beta}R_{j\beta} ,
\end{equation}  
where $R_{j\beta}$ is the coordinates of atom $j$ in the reference configuration. Then, the stress in Eq. (\ref{eq: pk stress 1}) can be written as
\begin{equation}\label{eq: pk stress 2}
	P_{\alpha\beta} = \frac{1}{V_0} \sum_{i=1}^{N} \sum_{m=1}^{M} \sum_{j \in \text{NB}_i} \frac{\partial E_i}{\partial G_{im}} \frac{\partial G_{im} } {\partial r_{j \alpha}} R_{j\beta}.
\end{equation}
Cauchy stress can be further calculated by
\begin{equation} \label{eq: cauchy stress from PK}
	\sigma_{\alpha\gamma} = \det ({\boldsymbol F})^{-1} \sum_{\beta=1}^{3}P_{\alpha\beta} F_{\gamma \beta}.
\end{equation}
Substitute Eq. (\ref{eq: pk stress 2}) into Eq. (\ref{eq: cauchy stress from PK}), and then apply Eq. (\ref{eq: coordinate mapping}) and $V = V_0 \det ({\boldsymbol F}) $, which is the volume in current configuration, we can get 
\begin{equation} \label{eq: cauchy stress}
	\sigma_{\alpha\beta} = \frac{1}{V}  \sum_{i=1}^{N} \sum_{m=1}^{M} \sum_{j \in \text{NB}_i} \frac{\partial E_i}{\partial G_{im}} \frac{\partial G_{im} } {\partial r_{j \alpha}} r_{j\beta}
\end{equation}
after replacing $\gamma$ with $\beta$.

\bibliographystyle{elsarticle-num} 
\bibliography{ref}





% \end{thebibliography}

\end{document}